\documentclass[aps,prl,reprint,groupedaddress,footinbib]{revtex4-1}
\usepackage{amsmath}
\usepackage{graphicx}
\usepackage{verbatim}
\usepackage{bbm}
\usepackage{color}

\begin{document}

	\title{Enhancement of the upper critical field\\ in disordered transition metal dichalcogenide monolayers }

	\author{Stefan Ili\'{c}, Julia S. Meyer, and Manuel Houzet}
	
	\affiliation{\textcolor{black}{Univ.~Grenoble Alpes, CEA, INAC-PHELIQS, F-38000 Grenoble, France}}

	\begin{abstract}
		We calculate the effect of impurities on the superconducting phase diagram of transition metal dichalcogenide monolayers in the presence of an in-plane magnetic field. 
		Due to strong intrinsic spin-orbit coupling, the upper critical field greatly surpasses the Pauli limit at low temperatures. We find that it is insensitive to intravalley scattering and, ultimately, limited by intervalley scattering. 
	\end{abstract}
	
	\maketitle
	
	\emph{Introduction.-} Transition metal dichalcogenide (TMDC) monolayers are recently discovered two-dimensional (2D) semiconductors of the form $\text{MX}_2$ (M=Mo, Nb,  or W, and X=S, Se, or Te), with  a hexagonal lattice structure similar to graphene, but with two inequivalent sites in the unit cell~\cite{review1-TMDC,review2-TMDC}. Like graphene, these materials exhibit a valley degree of freedom and have minima/maxima of conduction/valence bands at the corners $\mathbf{K}$ and $-\mathbf{K}$ of the Brillouin zone~\cite{mak}.  Unlike graphene, however,  the absence of inversion symmetry allows for a large, direct band gap, making them promising candidates for a new generation of transistors~\cite{radisavljevic,review1-TMDC}. 
	
	Due to the heavy constituent atoms, TMDC  monolayers exhibit a very large intrinsic spin-orbit coupling (SOC), often called Ising SOC \cite{zhou}, which acts as an effective Zeeman field perpendicular to the plane of the material with opposite orientations in the two valleys \cite{zhu2011giant,kormanyos2015k,xiao}. As a consequence, a large valley-dependent spin-splitting occurs in the valence band as well as in the conduction band, though with a much smaller magnitude { (due to the predominant $d_{x^2-y^2}\pm id_{xy}$ and $d_{z^2}$ orbital character of carriers, respectively)}. Recent optical investigations have confirmed that electrons from different valleys can be excited selectively with circularly polarized light \cite{mak2012control,zeng2012valley}. These properties open the door for novel applications in spintronics and so-called valleytronics \cite{wang, xiao}.   
	
	The coupling between the spin and valley degrees of freedom has remarkable repercussions for the superconducting properties that have been reported in these materials.  In heavily $n$-doped ionic-gated $\text{MoS}_2$ flakes, a 2D superconducting phase with a critical temperature $T_c$ around 10 K has been observed~\cite{saito,lu}. Interestingly, the in-plane upper critical field $\textcolor{black}{H_{c2}}$ reaches up to 60 T, thus greatly surpassing the Pauli limit, $H_P=\sqrt 2{\Delta_0}/(g\mu_B)$, where $\Delta_0\simeq 1.76 k_BT_c$, $\mu_B$ is the Bohr magneton, and $g$ the $g$-factor. This is interpreted as a consequence of the interplay of Ising SOC and the 2D nature of the materials. Namely, 
	in 2D materials the orbital pair-breaking effect is largely suppressed for an in-plane magnetic field~\cite{tinkham}. On the other hand, due to the Ising SOC, the in-plane magnetic field is also not efficient for breaking Cooper pairs by the paramagnetic effect, as they are formed of electrons in opposite valleys with strongly pinned out-of-plane spins. Related results have also been reported in superconducting $\text{NbSe}_2$ monolayers in the $p$-doped regime~\cite{xi2015ising}. \textcolor{black}{Existing theories only considered the clean case \cite{saito,frigeri} and do not describe all the experimental observations.}
	
	In this work, we establish the theory of the upper critical field for Ising superconductors at arbitrary disorder strength, assuming a conventional $s$-wave Cooper pairing. The enhancement of the in-plane upper critical field above the Pauli limit is a general feature of superconductivity in the presence of SOC. It has been predicted in disordered 2D superconductors with spin-orbit scattering \cite{maki,klemm} or Rashba SOC \cite{barzykin2002}.
	The latter has been invoked to interpret recent experiments on oxide interfaces \cite{reyren2009}
	and Pb monolayer films \cite{sekihara2013}. \textcolor{black}{Ising SOC leads to qualitatively different effects. Using a simple model, we show that Ising superconductivity results in a much larger enhancement and exhibits several new and interesting properties. Namely, $H_{c2}$ diverges at low temperature -- a phenomenon which is robust to intravalley scattering. On the other hand, intervalley scattering provides an effective spin-flip scattering mechanism and leads to the saturation of $H_{c2}$, consistent with the experimental findings. Furthermore, we show that, in contrast with Rashba SOC \cite{edelstein}, Ising SOC does not stabilize a spatially non-uniform superconducting phase. }

	Understanding the role of impurities is important for future applications of TMDCs and their \textcolor{black}{incorporation into van der Waals heterostructures \cite{geim}}. The doping techniques used to prepare the superconducting samples, as well as the defects in the crystal lattice could be the source of significant disorder. Recent weak localization measurements in $\text{MoS}_2$ monolayers suggest substantial intervalley scattering, attributed to a high concentration of sulfur vacancies in the monolayer \cite{schmidt}. Our result will contribute to identifying the superconducting properties of these materials in the presence of disorder \textcolor{black}{and assess their potential for applications such as in superconducting spintronics}.

	\emph{The model.-} The Hamiltonian describing the normal state of TMDC monolayers in the vicinity of the $\pm\mathbf{K}$ points in the presence of a parallel magnetic field is \cite{kormanyos2015k} (we use units, where $\hbar=k_B=1$):
	\begin{eqnarray}
	\mathcal{H}_{\mathbf{q}}&=&v(q_x\sigma_x\eta_z +q_y\sigma_y)+E_g\sigma_z\nonumber \\
	&&+\textcolor{black}{[\Delta_{A}(1+\sigma_z) +\Delta_{B}(\sigma_z-1)]s_z\eta_z +h s_x}.
	\label{original}
	\end{eqnarray}
	Here, $\mathbf{q}=(q_x,q_y)$ is a small momentum measured from $\pm \mathbf{K}$, $v$ is the velocity associated with the linearized kinetic dispersion, $E_g$ is the difference in on-site energy responsible for the opening of a band gap, and $\Delta_{A}$ and $\Delta_{B}$ are spin-splitting parameters on two different sublattices.
	Furthermore, $\sigma_{x,y,z}$, $s_{x,y,z}$, and $\eta_{x,y,z}$ are Pauli matrices acting in sublattice, spin, and valley spaces, respectively, and the Zeeman energy $h=\frac12 g\mu_B B$ is related with the amplitude of the magnetic field and the in-plane $g$-factor.
	
	\begin{figure}[h]
		\includegraphics[width=0.4 \textwidth]{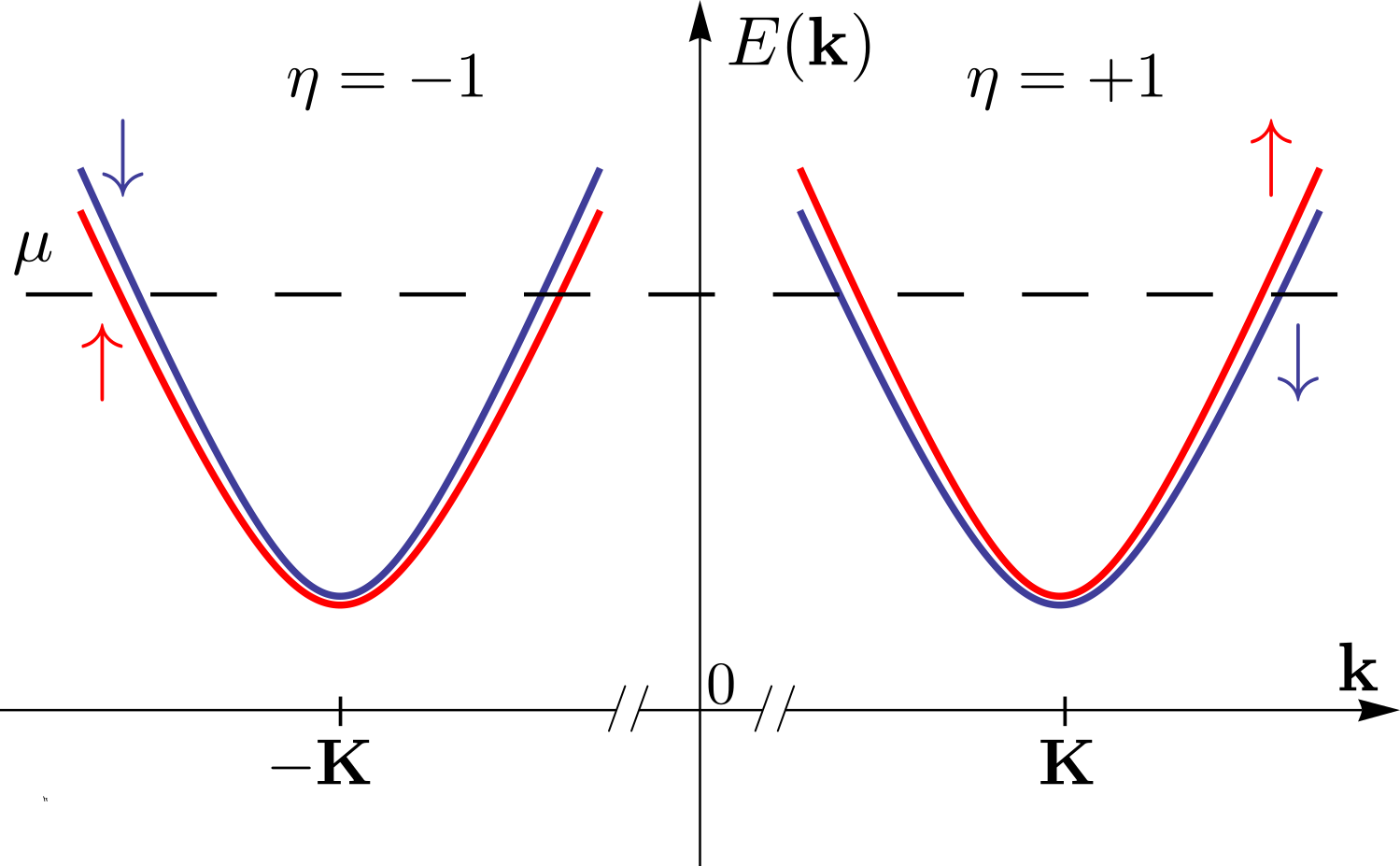}
		\caption{\label{fig1} Schematic representation of the conduction band of TMDC monolayers in the vicinity of 
			the corners of the Brillouin zone as described by Eq.~\eqref{eq:Hcb}. The spin splitting in the two valleys is opposite due to the so-called Ising SOC.}
	\end{figure}
	
	We proceed by projecting the Hamiltonian (\ref{original})  to the conduction band ($n$-doped regime). The spin-independent part of the Hamiltonian ~\eqref{original} gives the dominant contribution to the energy of the system and has a simple spectrum $\xi_{\mathbf{q}}=\sqrt{v^2|\mathbf{q}|^2+E_g^2}$. It is diagonalized by the unitary transformation $U_{\mathbf{q}}=\exp[i\alpha_{\mathbf{q}}\sigma_y\eta_z]\exp[i\beta_{\mathbf{q}}\sigma_z\eta_z]$, with $\tan (2\alpha_{\mathbf{q}})=v|\mathbf{q}|/E_g$ and $\tan (2\beta_{\mathbf{q}})=q_y/q_x$. 
	By assuming  $\xi_{\mathbf{q}}\gg \Delta_{A},\Delta_{B}$ and by projecting $U_{\mathbf{q}}\mathcal{H}_{\mathbf{q}}U_{\mathbf{q}}^\dagger$ onto the conduction  band, we obtain
	\begin{equation}
	\mathcal{H}_{\eta\mathbf{q}}=\xi_\mathbf{q}+\eta \Delta_{so}(\mathbf{q}) s_z+hs_x,
	\label{eq:Hcb}
	\end{equation}
	where $\eta=\pm1$ represents the valley index and $\Delta_{so}(\mathbf{q})=1/2 \textcolor{black}{[\Delta_{A}-\Delta_{B} + (\Delta_{A}+\Delta_{B})E_g/\xi_{\mathbf{q}}]}$ is an effective SOC parameter \footnote{Similar considerations in graphene, where $\Delta_{B}=\Delta_{A}$ and $E_g=0$, lead to the suppression of Ising SOC in the projected basis}. A Hamiltonian similar to Eq.~\eqref{eq:Hcb} with $\xi_{\mathbf{q}}\to-\xi_{\mathbf{q}}$ holds in the valence band ($p$-doped regime).
	Below, we will assume that the chemical potential $\mu$ is sufficiently far above $E_g$ \footnote{Ref.~\cite{sosenko2017} analyzes the effect of disorder on a distinct topological superconducting phase when $\mu$ lies in the vicinity of $E_g$.} on the relevant energy scales determining the superconducting properties ($h,\Delta,T,\dots \ll |\mu-E_g|$), so that $\Delta_{so}$ can be taken as a constant. The energy spectrum of Hamiltonian~\eqref{eq:HBCS} is illustrated in Fig.~\ref{fig1}.
	
	Superconductivity is described using a standard BCS Hamiltonian, where the singlet pairing of electrons into Cooper pairs is necessarily intervalley. The corresponding mean-field Hamiltonian reads
	\begin{equation}
	H_{BCS}=\sum_{\eta \mathbf{q}}c^\dagger_{\eta \mathbf{q} } \mathcal{H}_{\eta\mathbf{q}}  c_{\eta \mathbf{q}}
	+
	\Delta \sum_{\eta \mathbf{q}} c^\dagger_{\eta \mathbf{q} \uparrow} c^\dagger_{\bar{\eta}\mathbf{\bar{q}} \downarrow}+\text{h.c.},
	\label{eq:HBCS}
	\end{equation}
	where $c_{\eta \mathbf{q} }=(c_{\eta \mathbf{q} \uparrow},c_{\eta \mathbf{q} \downarrow})^T$ 
	is an annihilation operator for spin-up and spin-down electrons, and $\Delta$ is the superconducting order parameter. For compactness,  we use the abbreviations $\bar{\eta}=-\eta$ and $\bar{\mathbf{q}}=-\mathbf{q}$. In the vicinity of the second-order superconducting phase transition, $\Delta$ solves the linearized self-consistent gap equation
	\begin{equation}
	\Delta=\frac{\lambda T}{4}\sum_{\eta, \mathbf{q}, |\omega|<\Omega}\text{Tr}[i s_y \mathcal{G}_{\bar{\eta} \bar{\mathbf{q}}\omega}^+\Delta i s_y \mathcal{G}_{{\eta} {\mathbf{q}}\omega}^-],
	\label{selfcons}
	\end{equation}
	where $\lambda$ is the BCS pairing amplitude and $\Omega$ is a cut-off frequency. The particle and hole Green's functions are given as $\mathcal{G}^-_{\eta\mathbf{q}\omega}=(i\omega-\mathcal{H}_{\eta\mathbf{q}})^{-1}$ and 
	$\mathcal{G}^+_{\eta\mathbf{q}\omega}=(-i\omega-\mathcal{H}_{\eta\mathbf{q}}^T)^{-1}$, respectively, with Matsubara frequencies $\omega=(2n+1)\pi T$, $n\in\mathbbm{Z}$, at temperature $T$.

	\emph{Scattering potential.-}  
	The effect of impurities in 2D hexagonal lattices was extensively studied in the case of graphene \cite{aleiner2006effect,altland2006low}.  Here we study  the dominant scattering terms, namely spin-independent  intra- and intervalley scattering, which may originate from long-range Coulomb interactions or defects in the lattice. Their contribution, expressed in the same basis as the normal state Hamiltonian (\ref{original}), reads
	\begin{equation}
	\mathcal{H}_D(\mathbf{q-q'})=V_1(\mathbf{q-q'})+\sum_{i=x,y}V_{2i}(\mathbf{q-q'})\eta_i.
	\label{origdis}
	\end{equation}
	The random disorder potentials are characterized by the 
	Gaussian correlators $\langle V_{\alpha}(\mathbf{q})V_{\beta}(\mathbf{\bar{q}'}) \rangle=w_{\alpha}^2\delta_{\alpha,\beta}\delta_{\mathbf{q},\mathbf{q'}}$, where $\alpha,\beta=1,2x,2y$, and the brackets denote disorder averaging. The ratio $w_{2i}/w_{1}\sim 1/(|\mathbf{K}|^2 R^2)$, 
	which is related with the range $R$ of the potential created by a single impurity, 
	is typically small for remote impurities and of order 1 for lattice defects.

	Then, projecting the Hamiltonian $U_{\mathbf{q}}\mathcal{H}_D(\mathbf{q-q'})U^\dagger_{\mathbf{q'}}$ onto the conduction band, we find that Eq.~\eqref{eq:HBCS} has to be supplemented with the disorder term
	\begin{equation}
	H_D=\sum_{\eta \mathbf{q}  \mathbf{q'}}
	\mathcal{V}_{1\mathbf{qq'}}c^\dagger_{\eta \mathbf{q}}c_{\eta \mathbf{q'}}  
	+\mathcal{V}_{2\mathbf{qq'}}c^\dagger_{\eta\mathbf{q}}c_{\bar{\eta} \mathbf{q'}} +\text{h.c.},
	\label{projdis}
	\end{equation}
	where $\mathcal{V}_{1\mathbf{qq'}}=V_1(\mathbf{q-q'})\cos\frac{\theta-\theta'}{2}$ and $\mathcal{V}_{2\mathbf{qq'}}=[i V_{2x}(\mathbf{q-q'})+ V_{2y}(\mathbf{q-q'})]\sin\frac{\theta+\theta'}{2}$ 
	with ${\bf q}^{(')}=|{\bf q}^{(')}|(\cos\theta^{(')},\sin\theta^{(')})$. 
	
\begin{figure}[h!]
		\includegraphics[width=0.4 \textwidth]{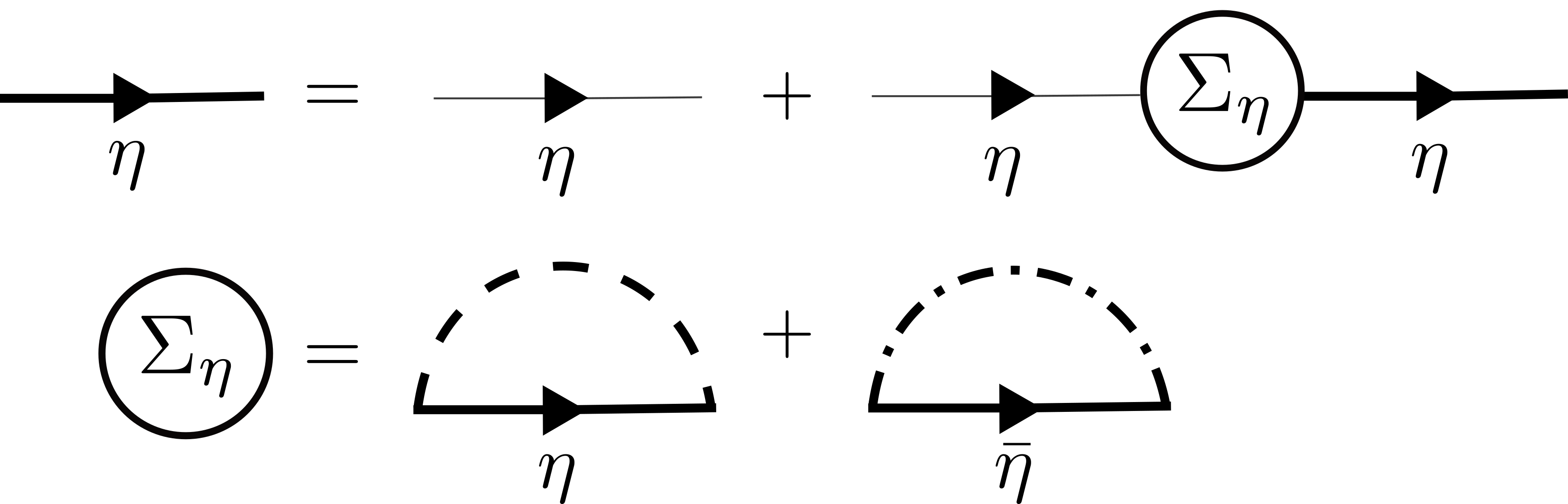}
		\caption{\label{sigma} Diagrammatic representation of the Dyson equation and the self-energy in the self-consistent Born approximation. The thin arrow represents the ``bare", disorder-free Green's function $\mathcal{G}_{\eta}^{\pm}$, while the thick arrow is the disorder-averaged Green's function $\langle\mathcal{G}_{\eta}^{\pm}\rangle$. The dashed and dot-dashed impurity lines represent intra- and intervalley scattering events, respectively.}
	\end{figure}

	The disorder-averaged Green's function can be calculated from the Dyson equation represented diagrammatically in Fig.~\ref{sigma}. That is, $\langle \mathcal{G}^{\pm}_{\eta \mathbf{q}\omega}\rangle=({\mathcal{G}^{\pm}_{\eta\mathbf{q}\omega}}^{-1}-\Sigma_{\eta}^{\pm})^{-1}$, where the self-energy $\Sigma_{\eta}^{\pm}$ is obtained using the self-consistent Born approximation. As a result, we find $\Sigma_\eta^{\pm}=\mp \textcolor{black}{i} [1/(2\tau_1)+1/(2\tau_2)]   \text{sgn}(\omega)$, where $1/\tau_1=\pi\nu w_1^2$ and $1/\tau_2=\pi\nu (w_{2x}^2+w_{2y}^2)$ are the intra- and intervalley elastic scattering rates, respectively, $\nu=\mu/(2\pi v_F^2)$ is the density of states per spin at the Fermi level, and $v_F$ is the Fermi velocity.
	
\begin{figure}[h!]
		\includegraphics[width=0.48\textwidth]{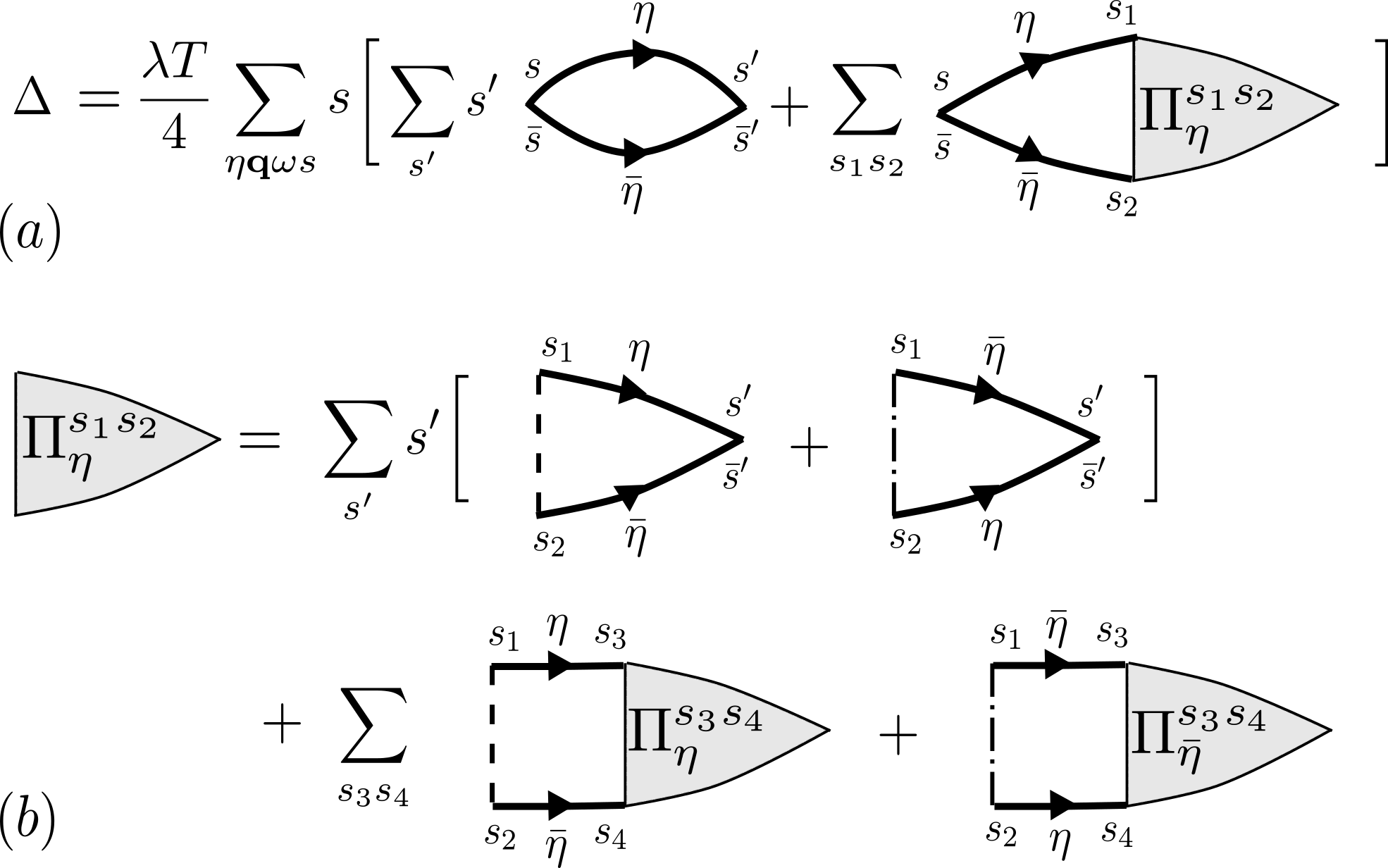}
		\caption{\label{diagrams}(a)  Diagrammatic representation of the disorder-averaged self-consistency condition given by Eq.~(\ref{selfcons}). (b) Bethe-Salpeter equation for the renormalized vertex functions $\Pi^{ss'}_{\eta}$. For the definition of diagram elements, see Fig.~\ref{sigma}. We use the abbreviation $\bar s=-s$.} 
	\end{figure}

	\emph{Upper critical line.-} The upper critical field \textcolor{black}{$h_{c2}(T)=\frac12 g \mu_B H_{c2}(T)$} 
	can be calculated from the disorder-averaged gap equation~(\ref{selfcons}). This involves finding the averages of the products of two Green's functions, which we do in the standard ladder approximation, as shown in Fig.~\ref{diagrams}(a). 
	The first diagram represents the bare vertex, while the second one is a ladder diagram, expressed  in terms of eight vertex functions $\Pi_\eta^{ss'}$
	(4 combinations of spin indices and 2 values of the valley index), which solve coupled Bethe-Salpeter equations, see Fig.~\ref{diagrams}(b).  
	Solving these equations yields a remarkably simple expression~\cite{SM}:  
	\begin{align}
	&\ln\frac{T}{T_c} \nonumber \\
	& =2\pi T \sum_{\omega>0}\left[\frac{\omega(\omega+\frac{1}{\tau_2})+\Delta_{so}^2}{(\omega+\frac{1}{\tau_2})(\omega^2+h_{c2}^2+\Delta_{so}^2)-\frac{\Delta_{so}^2}{\tau_2}}-\frac1\omega\right],
	\label{result}
	\end{align}
	where we used the standard BCS result $T_c \simeq\  1.13  \Omega  e^{-1/(\lambda\nu)}$.
	Note that Eq.~\eqref{result} holds for arbitrary values of the intravalley scattering rate. 
	Below we analyze this equation, which is the main result of this Letter.
	
	\emph{Without intervalley scattering.-} In the absence of intervalley scattering, $1/\tau_2=0$, the critical line given by Eq.~\eqref{result} does not depend on disorder (Anderson theorem). In that case, Eq.~\eqref{result}  can be alternatively expressed as
	\begin{equation}
	\ln\frac{T_{c}}{T}=\frac{h_{c2}^2}{h_{c2}^2+\Delta_{so}^2}\Re\bigg[\psi\bigg(\frac{1}{2}+\frac{i\sqrt{h_{c2}^2+\Delta_{so}^2}}{2\pi T}\bigg)-\psi\bigg(\frac{1}{2}\bigg)\bigg],
	\label{clean}
	\end{equation} 
	where $\psi(z)$ is the digamma function. In this form, it ressembles -- and generalizes to arbitrary disorder -- an expression derived by Frigeri \emph{et al.}~\cite{frigeri} in the clean case.
	It also reproduces the results of Ref.~\cite{saito}, where the linearized gap equation was solved numerically in the disorder-free case, using a complex multi-band model.
	
	The result can be understood as follows. The effective magnetic field in the two valleys is given by $\textcolor{black}{\mathbf{h}}^{\rm eff}_\eta=h{\bf e}_x+\eta\Delta_{so}{\bf e}_z$, where ${\bf e}_i$ is a unit vector in the $i$-direction. Electrons that are both aligned or anti-aligned with their respective \lq\lq local\rq\rq\ field have the same energy when their momenta are opposite and, thus, their contribution to pairing is not affected by the field. As at finite $h$ the local fields are not along the same axis, however, they enter the gap equation with a suppressed weight $\Delta_{so}^2/(h^2+\Delta_{so}^2)$, determined by the overlap of their spin directions. If one electron is aligned whereas the other electron is anti-aligned with their respective \lq\lq local\rq\rq\ field, they have an energy difference of $\textcolor{black}{2}\sqrt{h^2+\Delta_{so}^2}$ when their momenta are opposite and, thus, their contribution to pairing is suppressed by the field. Here the weight is given as $h^2/(h^2+\Delta_{so}^2)$, which is what appears in Eq.~\eqref{clean}. Since intravalley scattering does not allow for spin flips, it does not change the result.
	\begin{figure}[h!]
		\includegraphics[width=0.48 \textwidth]{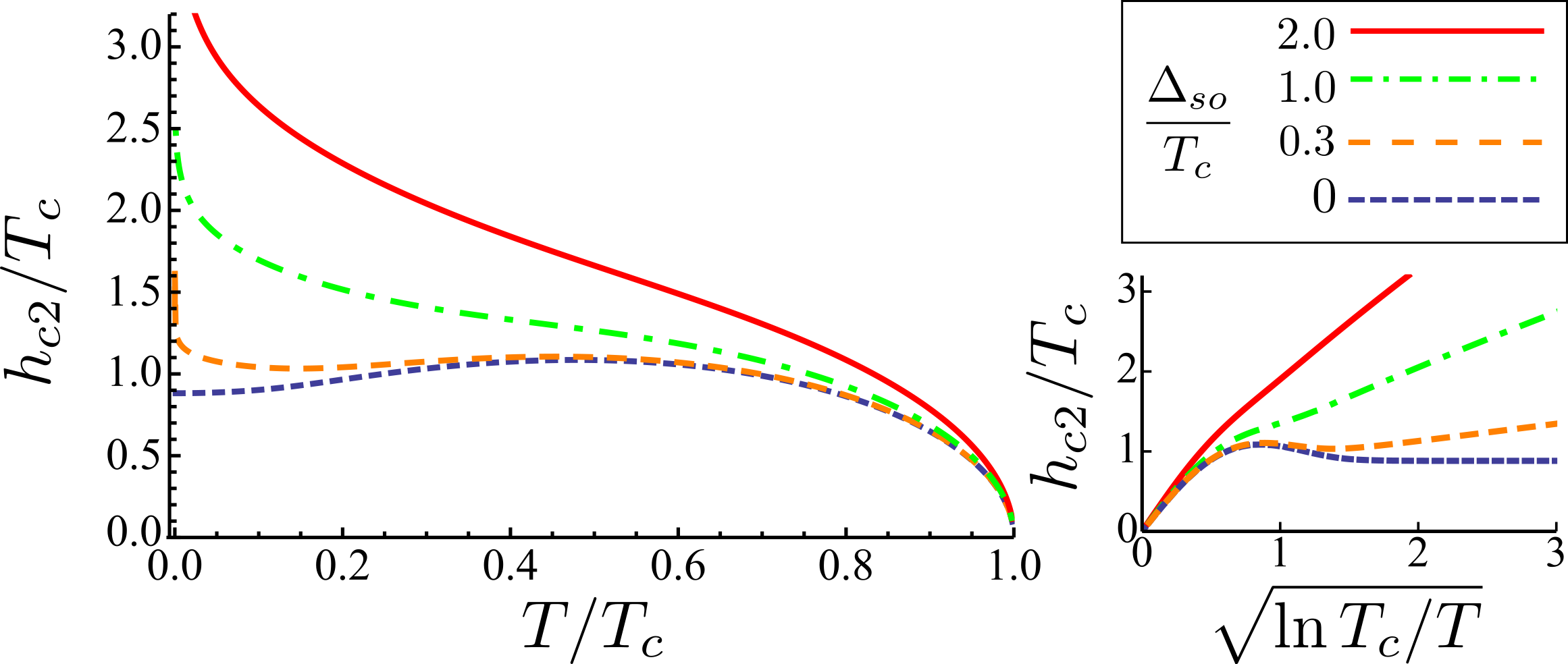}
		\caption{\label{fig4} Upper critical field as a function of temperature in the absence of intervalley scattering for different values of Ising SOC, as described by Eq.~\eqref{clean}. The plot on the right shows the same result but with a different scale for the $x$-axis to illustrate the logarithmic divergence at low temperature when $\Delta_{so}\neq0$.}
	\end{figure}

	\begin{figure*}[t!]
		\includegraphics[width=\textwidth]{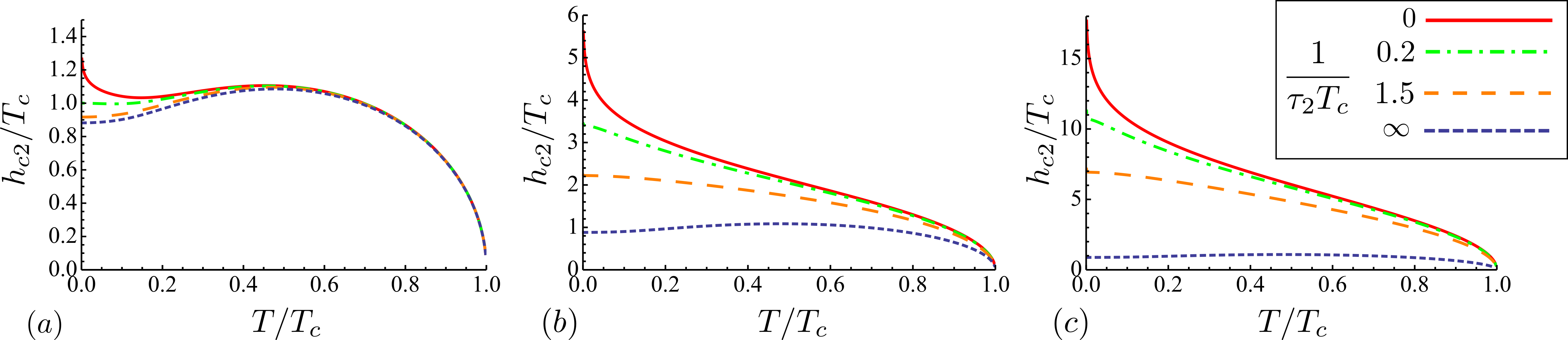}
		\caption{\label{graphs} Upper critical field as a function of the temperature for various strengths of Ising SOC and intervalley scattering: (a) $\Delta_{so}/T_c=0.3$, (b) $\Delta_{so}/T_c=3$, and (c) $\Delta_{so}/T_c=12$. The choice of parameters $\Delta_{so}/T_c=12$ and $1/(\tau_2 T_c)\textcolor{black}{=}1.5$ [dashed line in  (c)] gives a good fit of the experimental data from Ref.~\cite{saito} taking the $g$-factor to be $g=2$.}
	\end{figure*}
	As seen in Fig.~\ref{fig4}, $h_{c2}$ is enhanced in the presence of spin-orbit coupling, especially at low temperatures.  In fact, it diverges in the zero-temperature limit for finite $\Delta_{so}$. Physically, this can be understood as a consequence of the inability of the Zeeman field to completely align the electron spins in the in-plane orientation, due to the anti-parallel out-of-plane field provided by the Ising SOC.

	For weak Ising SOC ($\Delta_{so} \ll \Delta_0 $), the critical curve $h_{c2}(T)$ significantly deviates from the conventional one only at very low temperature, 
	where it diverges (in logarithmic accuracy) as  
	\begin{equation}
	h_{c2}\propto \Delta_{so}\sqrt{\ln\frac{T_c}{T}} \qquad\text{for} \qquad \frac{T}{T_c}\ll \exp \bigg(-c \frac{\Delta_{0}^2}{\Delta_{so}^2}\bigg),
	\label{weakSOC}
	\end{equation}
	where $c$ is a constant of order 1. Close to $T_c$, one obtains the standard result $h_{c2}\approx 2.16 T_c\sqrt{1-T/T_c}$.

	In the more interesting case of large Ising SOC, $\Delta_{so}\gg \Delta_0$, Eq.~\eqref{clean} yields a logarithmic divergence starting at higher temperatures,
	\begin{equation}
	h_{c2}
	\approx\Delta_{so}\sqrt{\ln\frac{T_{c}}{T}/\ln\frac{2\Delta_{so}}{\Delta_0}}\qquad\text{for}\qquad \frac{T}{T_c}\ll \frac{\Delta_0}{\Delta_{so}}.
	\label{logdiv}
	\end{equation}
	Close to $T_{c}$, the critical field exhibits a standard square-root dependence on temperature, but with an enhanced prefactor, 
	\begin{equation}
	h_{c2}
	\approx \Delta_{so}\frac1{\sqrt{\ln \frac{2\Delta_{so}}{\Delta_0}}}\sqrt{1-\frac{T}{T_{c}}}.
	\label{ctemp}
	\end{equation}

	\emph{With intervalley scattering.-}    
	At finite magnetic field, intervalley scattering provides an effective spin-flip mechanism, since electrons scattered between two valleys \lq\lq feel\rq\rq\ opposite values of the Ising SOC field. This pair-breaking effect leads to a saturation of $h_{c2}$ at zero temperature, 
	as illustrated in Fig.~\ref{graphs}.  
	For weak intervalley disorder, ${1}/{\tau_2}\ll \Delta_0\ll \Delta_{so}$, we estimate the zero-temperature critical field as
	\begin{equation}
	h_{c2}\approx \Delta_{so}\sqrt{\ln( \Delta_{0}\tau_2)/\ln\frac{2\Delta_{so}}{\Delta_0}}
	\label{tempdis}
	\end{equation}
	in logarithmic accuracy. In the vicinity of $T_{c}$, the critical line is still described by Eq.~(\ref{ctemp}) in that parameter regime.

	On the other hand, the standard \textcolor{black}{expression for the second-order} paramagnetically-limited critical line, given by Eq.~\eqref{clean} at $\Delta_{so}=0$, is recovered at large disorder strength, $1/\tau_2\gg \Delta_{so}^2/\Delta_0$.
	In this regime, electrons are frequently scattered between two valleys and do not \lq\lq feel\rq\rq\ the effect of valley-dependent Ising SOC anymore.

	\textcolor{black}{{\em Nature of the transition.-} In the absence of SOC, the phase transition is a second-order transition into a uniform superconducting
		state only at sufficiently high temperature. A non-uniform (Fulde-Ferrell-Larkin-Ovchinnikov, or FFLO) phase could possibly contribute to the enhancement of $h_{c2}$, as it was recently discussed in
		clean bilayer TMDC superconductors \cite{FFLO}. In order to study the nature of the transition in the clean case, we generalize the expression for the gap equation~\eqref{selfcons} by adding quadratic corrections in a finite modulation wavevector and cubic corrections in the gap amplitude $\Delta$ \cite{SM}. We find that both do not affect the transition when $\Delta_{so}\gtrsim \Delta_0$. Moderate disorder is not expected to change these conclusions \cite{FN}.}

	{\em Discussion and conclusion.-} Experiments \cite{saito,lu,xi2015ising} revealed superconductivity in TMDC well above the Pauli limit. The measured fields remained, however, below the values expected in the clean case with Ising SOC only. In Ref.~\cite{saito}, Rashba SOC was considered as a possible mechanism for the suppression of $h_{c2}$ at low temperature.
	However, the model required an irrealistically large amplitude for the Rashba SOC. Our work shows that moderate intervalley scattering, which has already been invoked in Ref.~\cite{schmidt} in the normal state, could provide an alternative scenario for the saturation. The curve corresponding to $1/(\tau_2T_c)=1.5$ shown in Fig.~\ref{graphs}(c) gives a good fit of the experimental data from Ref.~\cite{saito} \textcolor{black}{using their estimate for $\Delta_{so}/T_c$}.

	In conclusion, we have studied the effect of disorder on TMDC monolayer superconductors. We have predicted that the large enhancement of the upper critical magnetic field is robust to intravalley scattering.
	Furthermore, we have identified intervalley scattering as a likely mechanism for the more moderate enhancement of $h_{c2}$ observed in experiment.  
	\textcolor{black}{Interestingly, TMDCs have been identified as a possible
		platform for topological superconductivity and Majorana
		fermions \cite{zhou,sosenko2017} provided that unconventional pairing
		takes place. The role of disorder within these scenarios can be investigated 
		within the theory frame provided by our work.}

	\begin{acknowledgments}
		We acknowledge funding from the Laboratoire d'excellence LANEF in Grenoble (ANR-10-LABX-51-01) and by the ANR through the grant  ANR-16-CE30-0019. \end{acknowledgments}

\onecolumngrid
\pagebreak
\clearpage

\setcounter{equation}{0}
\setcounter{figure}{0}
\setcounter{table}{0}
\setcounter{page}{1}
\renewcommand{\thefigure}{S\arabic{figure}}
\renewcommand{\theequation}{S\arabic{equation}}
\begin{center}
\textbf{\large Supplemental Material for ``Enhancement of the upper critical field\\ in disordered transition metal dichalcogenide monolayers" }
\end{center}

This supplemental material contains technical details on the derivation and analysis of the upper critical field $h_{c2}$ in transition metal dichalcogenide monolayers that were omitted in the main text.

In Sec.~I, we provide more information on the evaluation of the equations shown in diagrammatic form in Fig.~\ref{diagrams} in the main text, used to obtain Eq.~\eqref{result}. In Sec.~II, we provide the derivation of asymptotic formulas for the dependence of $h_{c2}$ on temperature $T$, when $T\rightarrow 0$ and $T\rightarrow T_c$, given by Eqs.~(\ref{weakSOC})-(\ref{tempdis}) in the main text, and we compare them with the numerical evaluation of $h_{c2}(T)$. 
In Sec.~III, we give evidence that the transition is indeed a second-order transition into a homogeneous state, as assumed in the main text, if spin-orbit coupling is sufficiently large, $\Delta_{so} \gtrsim 0.52 \Delta_0$. In particular, in the clean case, we rule out a possible spatially modulated (Fulde-Ferrell-Larkin-Ovchinnikov, or FFLO) state as well as a first-order transition by generalizing the gap equation~\eqref{selfcons}.

\section{I.  Evaluation of diagrams}
\label{sec1}
\begin{figure}[h!]
	\includegraphics[width=0.9 \textwidth]{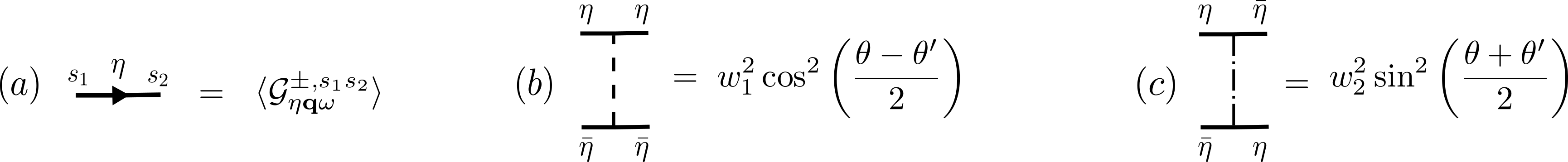}
	\caption{Feynman rules. $(a)$ Disorder-averaged Green's function. (b) Impurity line for intravalley scattering. (c) Impurity line for intervalley scattering. $\theta$ and $\theta'$ are polar angles associated with the momenta before and after the scattering event, as introduced in Eq.~\eqref{projdis}, and $w_2^2=w_{2x}^2+w_{2y}^2$. }
	\label{feyn}
\end{figure}

In Fig. \ref{feyn}, we show the Feynman rules for the ladder diagrams used in the main text.
The upper and lower branch in all the ladder diagrams represent the Green's functions $\mathcal{G}^+$ and $\mathcal{G}^-$, respectively, and all internal momenta (in-between scattering events) are integrated over. 

The diagrams for the disorder-averaged self-consistent gap equation, Fig.~\ref{diagrams}(a) in the main text, translate to
\begin{equation}
1=\frac{\lambda}{4}T\int\frac{d^2\,\mathbf{q}}{(2\pi)^2} \sum_{\eta\omega s s' s''}\ s \langle\mathcal{G}_{\eta\mathbf{q}\omega}^{+,ss'}\rangle \langle \mathcal{G}_{\bar{\eta}\mathbf{\bar{q}}\omega}^{-,\bar{s}s''}\rangle
\bigg[s'\delta_{\bar{s}'s''}+
\Pi_{\eta}^{s's''}(\theta)
\bigg]. 
\label{gapeqn}
\end{equation}
We evaluate the integrals over the momenta using the residue theorem with the substitution $\int \frac{d^2 \mathbf{q}}{(2\pi)^2}\approx \frac{\nu}{2\pi} \int d\xi_{\mathbf{q}}\int d\theta $, valid in the regime of chemical potentials specified in the main text.  In particular, we obtain
\begin{eqnarray}
\label{integrals}
\nu\int d\, \xi_{\mathbf{q}}
\langle \mathcal{G}_{\eta\mathbf{q}\omega}^{+,s_1 s_2}\rangle \langle \mathcal{G}_{\bar{\eta}\mathbf{\bar{q}}\omega}^{-,s_3 s_4}\rangle &=&\frac{\pi\nu\,\text{sgn}(\omega)}{2\tilde{\omega}(\tilde{\omega}^2+h^2+\Delta_{so}^2)}\bigg[
h^2(1-\delta_{s_1s_2})(1-\delta_{s_3s_4})+h(i\tilde{\omega}-s_3\eta\Delta_{so})(1-\delta_{s_1s_2})\delta_{s_4s_4}
\\
&&-h(i\tilde{\omega}-s_1\eta\Delta_{so})\delta_{s_1s_2}(1-\delta_{s_3s_4})+
(2\tilde{\omega}^2+h^2+\Delta_{so}^2\delta_{s_1\bar{s}_3}+s_12i\tilde{\omega}\eta\Delta_{so}\delta_{s_1s_3})\delta_{s_1s_2}\delta_{s_3s_4}
\bigg],
\nonumber
\end{eqnarray}
where we used the notation $\tilde{\omega}=\omega+{1}/({2\tau_1})+{1}/({2\tau_2}$).

We note that, in general, the renormalized vertex function $\Pi_{\eta}$ is dependent on the polar angle $\theta$, due to the anisotropy of the projected disorder potential, Eq.~\eqref{projdis}. The $\Pi_{\eta}$ are determined by a system of Bethe-Salpeter equations [corresponding to diagrams in Fig.~\ref{diagrams}(b) in the main text]:
\begin{eqnarray}
\label{bethesalp}
\Pi_\eta^{s_1 s_2}(\theta)=\int\frac{d^2\, \mathbf{q'}}{(2\pi)^2}\sum_{ss'}\bigg[&&
w_1^2\cos^2\bigg(\frac{\theta-\theta'}{2}\bigg) \langle \mathcal{G}_{\eta\mathbf{q'}\omega}^{+,s_1s}\rangle \langle \mathcal{G}_{\bar{\eta}\mathbf{\bar{q}'}\omega}^{-,s_2 s'}\rangle[s\delta_{\bar{s}s'}+\Pi_{\eta}^{ss'}(\theta)] \nonumber \\
&&+w_2^2\sin^2\bigg(\frac{\theta+\theta'}{2}\bigg) \langle\mathcal{G}_{\bar{\eta}\mathbf{\bar{q}'}\omega}^{+,s_1s}\rangle \langle \mathcal{G}_{\eta\mathbf{q'}\omega}^{-,s_2s'}\rangle
[s\delta_{\bar{s}s'}+\Pi_{\bar{\eta}}^{ss'}(\theta)]
\bigg].
\end{eqnarray}
We readily check that $\Pi_\eta(\theta)$ is fully determined by its first harmonics in $\theta$:
\begin{equation}
\Pi_{\eta}^{s_1 s_2}(\theta)=\Pi_{\eta 0}^{s_1 s_2}+\Pi_{\eta 1}^{s_1 s_2}\cos\theta+\Pi_{\eta 2}^{s_1 s_2}\sin\theta.
\label{pitheta}
\end{equation} 
Namely, by combining Eqs.~(\ref{pitheta}) and (\ref{bethesalp}), after integration over $\theta'$, we verify that no higher harmonics are generated. Furthermore, the equation for the constant part of the vertex functions $\Pi_{\eta 0}$ is decoupled from the angle-dependent parts $\Pi_{\eta 1}$ and $\Pi_{\eta 2}$. Replacing the $\Pi_{\eta}(\theta)$  in Eq.~(\ref{gapeqn}), we see that the angle-dependent contributions vanish after the integration over momenta. Therefore, it is sufficient to compute only $\Pi_{\eta 0}$.

The eight different $\Pi_{\eta 0}$ are determined from the linear system of equations obtained after integrating (\ref{bethesalp}) over angles:  
\begin{align}
\label{pizero}
\Pi_{\eta 0}^{s_1 s_2}=\int\frac{d\,\xi_{\mathbf{q}'} }{2\pi}\sum_{ss'}\bigg[
\frac{1}{\tau_1} \langle \mathcal{G}_{\eta\mathbf{q'}\omega}^{+,s_1s}\rangle& \langle \mathcal{G}_{\bar{\eta}\mathbf{\bar{q}'}\omega}^{-,s_2 s'}\rangle[s\delta_{\bar{s}s'}+\Pi_{\eta0}^{ss'}]
+\frac{1}{\tau_2} \langle\mathcal{G}_{\bar{\eta}\mathbf{\bar{q}'}\omega}^{+,s_1s}\rangle \langle \mathcal{G}_{\eta\mathbf{q'}\omega}^{-,s_2s'}\rangle
[s\delta_{\bar{s}s'}+\Pi_{\bar{\eta}0}^{ss'}]
\bigg].
\end{align}
Inserting the vertex functions that solve Eq.~\eqref{pizero} into Eq.~\eqref{gapeqn} yields Eq.~\eqref{result} in the main text.

\section{II.  Limiting cases}
\label{sec2}
In the following, starting from the full expression for the critical curve $h_{c2}(T)$, given by Eqs.~\eqref{result} and \eqref{clean} in the main text, we derive simple analytical expressions in the zero-temperature limit as well as close to the transition temperature $T_c$, for various strengths of the intervalley scattering rate. We compare them with the results of the numerical calculations of $h_{c2}$ for arbitrary disorder strength in the same temperature regimes.

\subsection{A.  Limit $T\rightarrow 0$}
In the absence of intervalley scattering, $1/\tau_2=0$, the critical curve $h_{c2}(T)$ is described by Eq. (\ref{clean}) in the main text. At low temperatures, using the asymptotic behavior of the digamma function $\psi(z)\approx \ln |z|$ for $|z|\gg 1$, Eq.~\eqref{clean} in the main text thus yields
\begin{equation}\label{eqs70}
\ln \frac{T_c }{T}=\frac{h_{c2}^2}{\rho^2}\ln \frac{4 e^\gamma \rho}{2\pi T},
\end{equation}
where $\gamma\approx 0.577$ 
and we introduced $\rho=\sqrt{h_{c2}^2+\Delta_{so}^2}$. In the limit $T\to0$, $h_{c2}$ diverges, as illustrated in Fig.~\ref{fig4} in the main text. Thus, we can approximate our results assuming $h_{c2}\gg \Delta_{so},\Delta_0$. Then, Eq.~\eqref{eqs70} yields
\begin{equation}
\ln \frac{2h_{c2}}{\Delta_0}\approx \frac{\Delta_{so}^2}{h_{c2}^2}\ln \frac{T_c}{T}.
\label{eqs7}
\end{equation}
In the limit of strong Ising SOC, $\Delta_{so}\gg\Delta_0$, the condition $h_{c2}\gg\Delta_{so}$  allows one to simplify Eq.~\eqref{eqs7} in the temperature regime $T/T_c<\Delta_0/\Delta_{so}$ to Eq.~\eqref{logdiv} in the main text, in logarithmic accuracy. In the limit of weak Ising SOC, $\Delta_{so}\ll \Delta_0$, we see that the condition $h_{c2}\gg\Delta_{0}$ 
is fulfilled for $(\Delta_{so}/\Delta_0)^2\ln(T_c/T)\gg 1$.
Then, we are able to give a rough estimate provided by Eq.~\eqref{weakSOC} in the main text. 

At weak disorder, ${1}/{\tau_2}\ll \Delta_{so},\Delta_0$, the poles in the $\omega$-dependent terms in Eq. (\ref{result}) in the main text can be evaluated perturbatively to yield:
\begin{equation}
\ln\frac{T_{c}}{T}=\frac{\Delta_{so}^2}{\rho^2}\bigg[\psi\bigg(\frac{1}{2}+\frac{h_{c2}^2}{2\pi\tau_2\rho^2T}\bigg)-\psi\bigg(\frac{1}{2}\bigg)\bigg]+
\frac{h_{c2}^2}{\rho^2}\Re\bigg[\psi\bigg(\frac{1}{2}+\frac{i\rho}{2\pi T}\bigg)-\psi\bigg(\frac{1}{2}\bigg)\bigg].
\label{weakt2}
\end{equation} 
By comparing with Eq.~\eqref{clean} in the main text, we see that the main effect of weak intervalley scattering is to provide an effective pair-breaking rate for pairs of electrons that are both aligned or anti-aligned with their respective local fields, which yields an additional mechanism for the suppression of $h_{c2}$. As a consequence, $h_{c2}$ now saturates at $T\to 0$.
Assuming $1/\tau_2\ll \Delta_0\ll\Delta_{so}$ such that $h_{c2}\gg \Delta_{so}$ still holds, we find from Eq.~\eqref{weakt2} that
\begin{equation}
\ln \frac{2h_{c2}}{\Delta_0}\approx \frac{\Delta_{so}^2}{h_{c2}^2}\ln \tau_2h_{c2},
\end{equation}
which evaluates to Eq.~\eqref{tempdis} in the main text, in logarithmic accuracy.
For weak Ising SOC, $\frac{1}{\tau_2}\ll \Delta_{so}\ll\Delta_{0}$, similar estimates for $h_{c2}$ can be made up to a constant factor, in the same manner as in Eq.~\eqref{weakSOC}.

For large disorder strength, Eq.~\eqref{result} in the main text evaluates to
\begin{equation}
\ln \frac{T}{T_c}=2\pi T \sum_{\omega>0}\bigg[\frac{\omega+\tau_2\Delta_{so}^2}{\omega^2+h_{c2}^2}-\frac{1}{\omega}
\bigg]=\Re\bigg[\psi\bigg(\frac{1}{2}\bigg)-\psi\bigg(\frac{1}{2}+\frac{i h_{c2}}{2\pi T}\bigg)\bigg]+\frac{\pi \Delta_{so}^2\tau_2}{2h_{c2}}\tanh\frac{h_{c2}}{2T},
\label{dif}
\end{equation}
provided that only Matsubara frequencies $\omega\lesssim h_{c2}\ll1/\tau_2$ contribute to the sum. At $T\to 0$, Eq.~\eqref{dif} simplifies to 
\begin{equation}
\ln\frac{2h_{c2}}{\Delta_0}=\frac{\pi \Delta_{so}^2\tau_2}{2h_{c2}}.
\end{equation}
Thus, at large Ising SOC, $\Delta_{so}\gg\Delta_0$, we find the standard paramagnetic limit $h_{c2}=\Delta_0/2$ for the second-order phase transition if $1/\tau_2\gg\Delta_{so}^2/\Delta_0$. At weaker disorder, $\Delta_{so}\ll 1/\tau_2\ll\Delta_{so}^2/\Delta_0$, we obtain  $h_{c2}=\pi\tau_2\Delta^2_{so}/[2\ln(\tau_2\Delta_{so}^2/\Delta_0)]$. Further decreasing disorder, the condition  $h_{c2}\ll1/\tau_2$ breaks down and Eq.~\eqref{dif} cannot be used anymore.

\begin{figure}[t!]
	\includegraphics[width=\textwidth]{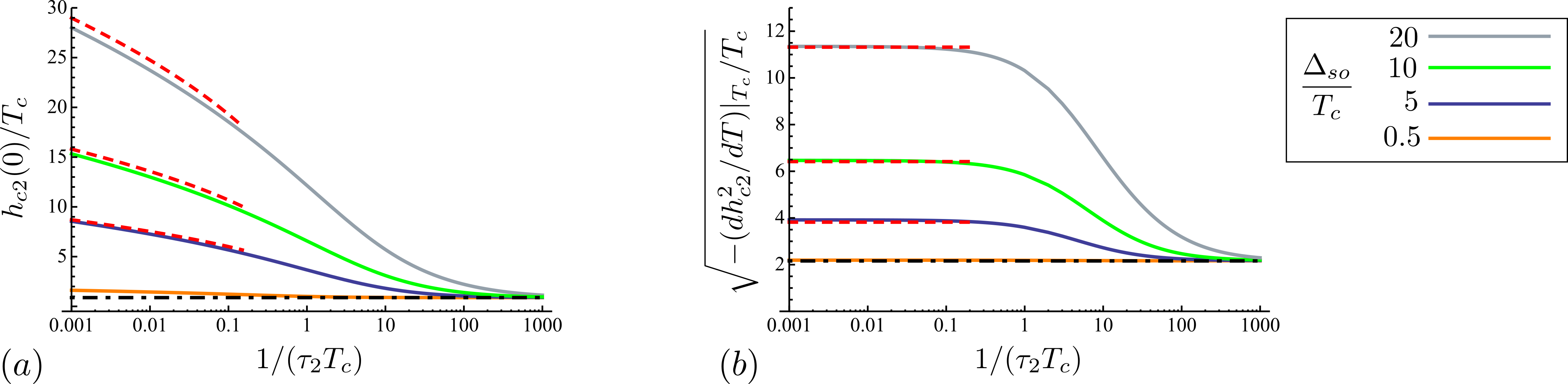}
	\caption{Behavior of the critical field $h_{c2}(T)$ close to $0$ and $T_c$ as a function of disorder strength for various values of the Ising SOC. We show numerical (solid lines) and approximate (dashed lines) results. (a) The critical fields $h_{c2}(0)$ at $T=0$, obtained from Eq.~\eqref{zerotemp} and from Eq.~\eqref{tempdis}, calculated up to the second order in the logarithmic approximation. The black dash-dotted line corresponds to the result in the absence of SOC, $h_{c2}=\Delta_0/2$. (b) Behavior of $h_{c2}$ close to $T_c$: we plot $1/\sqrt{C}=[-(dh^2_{c2}(T)/dT)|_{T_c}/T_c]^{1/2}$, obtained from Eq.~\eqref{eqs11} and from Eq.~\eqref{ctemp}. The black dash-dotted line corresponds to $1/\sqrt{C}=2\pi/\sqrt{7\zeta(3)}\approx 2.16$ in the absence of SOC. 
	}
	\label{figs2}
\end{figure}

To compare these results with the numerical evaluation of $h_{c2}(0)$, it is convenient to obtain the later from the following equation: 
\begin{equation}
\ln \frac{2 h_{c2}(0)}{\Delta_0}=\int_{0}^\infty d\omega \bigg[
\frac{\omega(\omega+\frac{1}{\tau_2})+\Delta_{so}^2}{(\omega+\frac{1}{\tau_2})(\omega^2+h_{c2}^2+\Delta_{so}^2)-\frac{\Delta_{so}^2}{\tau_2}}-\frac{\omega}{\omega^2+h_{c2}^2}
\bigg].
\label{zerotemp}
\end{equation}
We plot $h_{c2}(0)$ obtained this way in Fig.~\ref{figs2}(a) as a function of $1/\tau_2$. We verify that  Eq.~\eqref{tempdis} is in good agreement with these results in the relevant parameter regime and that $h_{c2}(0)$ reaches $\Delta_0/2$ for strong disorder. 

\subsection{B.  Limit $T\rightarrow T_c$}
Close to $T_c$, the critical field $h_{c2}$ is small, and the assumptions $\Delta_{so}\gg h_{c2}$ and $\ln \frac{T_c}{T}\approx \frac{T_c-T}{T_c}$ hold. Expanding Eq.~\eqref{result} in the main text in this regime yields a square-root behavior 
\begin{equation}
h_{c2}\simeq \sqrt{\frac{T_c(T_c-T)}{C}}
\end{equation}
with 
\begin{equation}
C= 2\pi T_c^3 \sum_{\omega>0} \frac{\omega+\frac{1}{\tau_2}}{\omega^2[\omega(\omega+\frac{1}{\tau_2})+\Delta_{so}^2]}
\label{eqs11}
\end{equation}
and $\omega=(2n+1)\pi T_c$ ($n$ integer).
Analyzing Eq.~\eqref{eqs11} in various regimes, we find
$C=7\zeta(3)/4\pi^2$ at weak disorder and weak Ising SOC, $1/\tau_2,\Delta_{so}\ll\Delta_0$, as well as at large disorder, $1/\tau_2\gg \Delta_{so}^2/\Delta_0\gg\Delta_{so}$. Outside those ranges of disorder, $C$ is suppressed -- and, thus, $h_{c2}$ is enhanced. For instance, $C=(T_c/\Delta_{so})^2\ln(2\Delta_{so}/\Delta_0)$ at weak disorder and large Ising SOC, $1/\tau_2\ll \Delta_0\ll\Delta_{so}$, and 
$C=\pi T_c/(4\tau_2\Delta_{so}^2)$  in an intermediate disorder range and at large Ising SOC,
$\Delta_0\ll 1/\tau_2,\Delta_{so}\ll \Delta_{so}^2/\Delta_0$.

We explore a wider window of disorder strengths numerically using Eq.~\eqref{eqs11}, as shown in Fig.~\ref{figs2}. The results match the approximate formulas close to $T_c$ given above at weak and strong disorder.

\section{III.  FFLO phase and the First-order phase transition}\label{sec3}

In the main text, we only considered the second-order transition into a homogeneous superconducting state.
In the following, we examine the conditions for realizing a FFLO phase and/or first-order phase transition in clean Ising superconductors. We show that both are absent for $\Delta_{so}\gtrsim 0.52 \Delta_0$. We do not expect moderate disorder, $1/\tau_2\ll {\rm max}(\Delta_0,\Delta_{so}^2/\Delta_0)$, to lead to a reappearance of the FFLO phase and/or first-order phase transition.

\subsection{A.  FFLO phase}
In the FFLO phase, the superconductor is spatially modulated. At the second-order transition into that state, we account for an exponentially modulated order parameter $\Delta( \mathbf{x})=\Delta_\mathbf{p}e^{i \mathbf{p}. \mathbf{x}}$, where  $\mathbf{p}$ is the modulation wavevector, by modifying the pairing term in the BCS Hamiltonian, 
\begin{equation}
H_{FFLO}=\sum_{\eta \mathbf{q}}c^\dagger_{\eta \mathbf{q} } \mathcal{H}_{\eta\mathbf{q}}  c_{\eta \mathbf{q}}
+\Delta_{\mathbf{p}} \sum_{\tau \mathbf{q}}c^{\dagger}_{\tau\mathbf{q}+\mathbf{p}/2\uparrow} c^{\dagger}_{\bar{\tau}\bar{\mathbf{q}}+\mathbf{p}/2\downarrow} +{\rm h.c.}.
\end{equation}
Then, in the vicinity of the phase transition, the amplitude of the order parameter should solve the linearized self-consistent gap equation
\begin{equation}
\Delta_{\mathbf{p}}=\frac{\lambda T}{4}\sum_{\eta, \mathbf{q}, |\omega|<\Omega}\text{Tr}[i s_y \mathcal{G}_{\bar{\eta} \bar{\mathbf{q}}+\frac{\mathbf{p}}{2}\omega}^+\Delta_{\mathbf{p}} i s_y \mathcal{G}_{{\eta} {\mathbf{q}}+\frac{\mathbf{p}}{2}\omega}^-].
\end{equation} 
In order to consider the instability toward an FFLO state along the upper ciritcal line $h_{c2}(T)$, we further assume that the modulation wavevector is small, $v_F|\mathbf{p}|\ll \Delta_0$. The small momentum shift $\mathbf{p}/2$ in the Green's functions can be accounted for by shifting the energy $\xi_{\mathbf{q}+\frac{\mathbf{p}}{2}}\approx \xi_\mathbf{q}+v_F\mathbf{p}. \hat{\mathbf{q}}/2$, where we introduced $\hat{\mathbf{q}}=\mathbf{q}/|\mathbf{q}|$. Then, integration over the momenta can be carried out using the results given in Eq.~\eqref{integrals} by taking $\tilde{\omega}\to\omega-iv_F\mathbf{p}. \hat{\mathbf{q}}/2$. After summing over Matsubara frequencies, we obtain 
\begin{equation}
\ln \frac{T}{T_{c}}=\int \frac{d\theta}{2\pi}\Re\bigg\{
\frac{\Delta_{so}^2}{\rho^2} \bigg[
\psi\bigg(\frac{1}{2}\bigg) 
-\psi\bigg(\frac{1}{2}-\frac{iv_F\mathbf{p}. \hat{\mathbf{q}}}{4\pi T}\bigg)\bigg]
+\frac{h_{c2}^2}{\rho^2} 
\bigg[
\psi\bigg(\frac{1}{2}\bigg) -
\psi\bigg(\frac{1}{2}+\frac{2i\rho-iv_F\mathbf{p}. \hat{\mathbf{q}}}{4\pi T}\bigg)
\bigg]
\bigg\}.
\end{equation}
Expanding the above expression in small $\mathbf{p}$ and integrating over angles yields
\begin{equation}
\ln \frac{T}{T_{c}}=-\frac{h_{c2}^2}{\rho^2}\Re \bigg[
\psi\bigg(\frac{1}{2}+\frac{i\rho}{2\pi T}\bigg)
-\psi\bigg(\frac{1}{2}\bigg)
\bigg]
+\frac{v_F^2|\mathbf{p}|^2T_c^2}{16\pi^2 h_{c2}^2T^2}F_1(T,\Delta_{so}),
\label{fflo}
\end{equation}
where
\begin{equation}
F_1(T,\Delta_{so})=-\left(\frac{h_{c2}}{T_c}\right)^2\left\{\psi^{(2)}\bigg(\frac{1}{2}\bigg)+\frac{h_{c2}^2}{\rho^2}\Re\bigg[ \psi^{(2)}\bigg(\frac{1}{2}+\frac{i\rho}{2\pi T}\bigg)-\psi^{(2)}\bigg(\frac{1}{2}\bigg)\bigg]\right\}.
\label{f1}
\end{equation}
Here,  $\psi^{(2)}(z)$ is the second derivative of the digamma function. 

The last term in Eq.~\eqref{fflo} is the correction to the result in the uniform case, Eq.~\eqref{clean}, due to the modulation. The instability toward the FFLO state is determined by the sign of $F_1$ along the line $h_{c2}(T)$ for the uniform state. Namely, if $F_1>0$ (resp. $F_1<0$), $h_{c2}$ decreases (resp. increases) when the order parameter is modulated.

We evaluate $F_1$ along the upper critical line derived for the uniform state in Fig.~\ref{figs3}(a). At $\Delta_{so}=0$, $F_1$ changes sign at $T^*=0.56 T_c$, signaling a transition into the FFLO state below that temperature. At small $\Delta_{so}$, we find that $F_1$ changes  sign at two temperatures $T^*_1$ and $T^*_2$, with 
$T^*_1<T<T^*_2$. The range of temperatures $T^*_1<T<T^*_2$, where the FFLO state can be expected shrinks as $\Delta_{so}$ increases, and it eventually disappears at $\Delta_{so}\gtrsim0.30 \Delta_0$, thus excluding the possibility of an FFLO phase at larger $\Delta_{so}$.

\subsection{B.  First-order phase transition}
In order to study the possibility of a first-order phase transition, the linearized self-consistency equation is not sufficient and we need to include higher order terms in $\Delta$. Thus, we write the self-consistent gap equation for the uniform phase [described by Eq.~\eqref{eq:HBCS}] up to third order in $\Delta$,
\begin{equation}
\Delta=\frac{\lambda T}{4}\sum_{\eta, \mathbf{q}, |\omega|<\Omega}\text{Tr}[i s_y \mathcal{G}_{\bar{\eta} \bar{\mathbf{q}}\omega}^+\Delta i s_y \mathcal{G}_{{\eta} {\mathbf{q}}\omega}^-+
i s_y \mathcal{G}_{\bar{\eta} \bar{\mathbf{q}}\omega}^+\Delta i s_y \mathcal{G}_{{\eta} {\mathbf{q}}\omega}^-\Delta i s_y \mathcal{G}_{\bar{\eta} \bar{\mathbf{q}}\omega}^+\Delta i s_y \mathcal{G}_{{\eta} {\mathbf{q}}\omega}^-].
\end{equation}
After evaluating the integrals over the products of four Green's functions using the residue theorem [in a similar fashion as in Eq.~\eqref{integrals}], and summing over Matsubara frequencies, we obtain 
\begin{equation}
\ln \frac{T}{T_{c}}=-\frac{h_{c2}^2}{\rho^2}\Re \bigg[
\psi\bigg(\frac{1}{2}+\frac{i\rho}{2\pi T}\bigg)
-\psi\bigg(\frac{1}{2}\bigg)
\bigg]+\frac{\Delta^2T_c^4}{16\pi h_{cs}^4T^2}F_2(T,\Delta_{so}),
\label{first}
\end{equation}
where 
\begin{equation} 
F_2(T,\Delta_{so})=
\frac{4h_{c2}^4}{\pi T_c^4}(2\pi T)^3\sum_{\omega>0}\frac{(\Delta_{so}^2+\omega^2)[h_{c2}^2(\Delta_{so}^2-3\omega^2)+(\Delta_{so}^2+\omega^2)^2]}
{\omega^3(h_{c2}^2+\Delta_{so}^2+\omega^2)^3}.
\label{f2}
\end{equation}
The last term in Eq.~\eqref{first} is the correction to the linearized gap equation~\eqref{clean} due to a finite amplitude of the order parameter in the vicinity of the transition. The order of the transition is determined by the sign of $F_2$ along the second-order transition line $h_{c2}(T)$. Namely, if $F_2>0$ (resp. $F_2<0$), the transition remains second-order (resp. a change of the order of the transition occurs).

\begin{figure}[t!]
	\includegraphics[width=\textwidth]{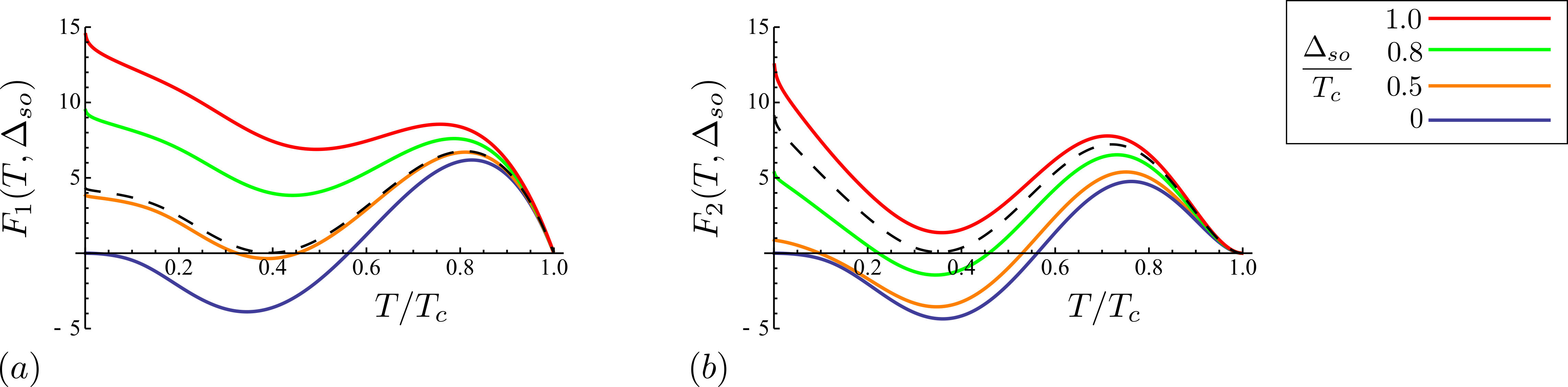}
	\caption{Temperature-dependence of $F_{1/2}(T,\Delta_{so})$ for various values of the Ising SOC. (a) FFLO: $F_1(T,\Delta_{so})$ defined in Eq.~\eqref{f1}, evaluated along the upper critical line in the uniform state. The dashed black line corresponds to the critical value of the Ising SOC, $\Delta_{so}^{FFLO}=0.53 T_c\simeq0.30\Delta_0$, above which the function $F_1$ remains positive for all temperatures. (b) First order phase transition: $F_2(T,\Delta_{so})$ defined in Eq.~\eqref{f2}, evaluated along the upper critical line in the uniform state. The dashed black line corresponds to the critical value of Ising SOC $\Delta_{so}^{1\to2}=0.92 T_c\simeq0.52\Delta_0$, above which the function $F_2$ remains positive for all temperatures. 
	}
	\label{figs3}
\end{figure}

At $\Delta_{so}=0$, we find that $F_2=(h_{c2}/T_c)^2F_1$. Thus, the sign change occurs at the same temperature and, as a consequence, for $T<T^*=0.56 T_c$  the transition into the FFLO state is in competition with a first-order transition. At finite $\Delta_{so}$, we evaluate $F_2$ along the upper critical line derived for the uniform state in Fig.~\ref{figs3}(b). We find that its temperature dependence is qualitatively similar to, though quantitatively different from $F_1$. Thus, a change of the transition order may occur in a finite temperature range, if Ising SOC is weak. On the other hand, $F_2$ remains positive at all temperatures if $\Delta_{so}\gtrsim0.52 \Delta_0$, and therefore the transition remains a second-order transition at larger Ising SOC.

\end{document}